# ML-Assisted Bulk Resource Allocation: Custom Outage-Based Loss Function and Reliability Analysis

Amir Masoud Molaei, *Senior Member, IEEE*, Nidhi Simmons, *Senior Member, IEEE*, David E. Simmons, Hien Quoc Ngo, *Fellow, IEEE*, and Okan Yurduseven, *Senior Member, IEEE*

***Abstract*—Machine learning (ML)-assisted outage-based resource allocation has recently emerged as an effective alternative to conventional scheduling methods in reliability-critical wireless systems. However, existing approaches are fundamentally limited to single-resource allocation, whereas modern and emerging systems increasingly require the simultaneous allocation of multiple resources to meet aggregate rate and reliability constraints. In this paper, we extend outage-based learning to the bulk resource allocation regime, where a user requires at least *D* reliable resources from a pool of *R* candidates. We first introduce a practical allocation policy, termed gate + top-*D* allocation (GTBA), which combines threshold-based admission control with ranking-based selection. We then propose a novel ranking-aware bulk outage loss (RBOL) that provides a differentiable surrogate for the bulk outage event induced by GTBA, explicitly accounting for both gate failures and ranking errors near the selection boundary. An exact reliability analysis is developed, establishing a decomposition of bulk outage probability (BOP), identifying dominant failure mechanisms and deriving an oracle lower bound that characterizes the fundamental performance limit. Extensive simulations under balanced, light and heavy stress regimes demonstrate that RBOL consistently outperforms conventional pointwise losses and baselines, achieving substantial reductions in BOP and remaining significantly closer to the oracle bound across a wide range of operating conditions. These results confirm that set-level ranking-aware training objectives are essential for reliable ML-assisted bulk resource allocation.**

## I. INTRODUCTION

MACHINE learning (ML) is increasingly adopted in wireless and Internet of Things (IoT) systems, including vehicular edge computing and task offloading, for prediction, channel estimation and resource allocation [1-4]. Traditional user scheduling and channel assignment mechanisms [5, 6] rely on deterministic rules or heuristics that require periodic and expensive estimation of channel quality indicators, but ML-assisted allocation [6, 7] (especially when paired with outage-based training objectives [8, 9]) has recently emerged as an attractive alternative. These methods enable a model to predict the likelihood that a given resource will fail to meet a target reliability threshold, allowing the system to allocate the safest resource without explicit channel decoding. Prior work in this direction has developed ML-assisted outage-based allocation strategies [9, 10], provided exact analytical outage expressions, including outage loss function (OLF), for single-resource allocation [9, 10], introduced custom loss functions that outperform conventional binary cross-entropy (BCE) in reliability-critical regimes [9, 11], and explored calibration methods to better align ML outputs with real-world reliability metrics [12]. Related ML-assisted resource allocation problems have also been studied in IoT edge/fog computing networks, including ML-based resource allocation for IoT edge computing [13], deep-reinforcement-learning-based multiuser resource control [14], and resource allocation for digital-twin-enabled vehicular edge computing [4].

Despite these advances, a fundamental limitation persists: all the above existing ML-assisted outage-based approaches assume that the system assigns exactly one resource per user. This assumption, while convenient analytically, is increasingly restrictive and fails to reflect the requirements of emerging sixth-generation (6G) architectures. Modern systems (including carrier aggregation [15], multi-beam multiple-input multiple-output [16], coordinated multi-point [17] and multi-band ultra-reliable low latency communications (URLLC) [18]) require allocating multiple resources to the same user, often to achieve an aggregate rate or to diversify against deep fades. In these scenarios, a user may need $D$ reliable resources out of $R$ available candidates, and is considered in outage not when a single resource fails, but when fewer than $D$ of the allocated resources satisfy the reliability threshold.

Extending ML-assisted outage-based allocation to this multi-resource (*bulk*) domain introduces several new challenges. First, while the ML model outputs per-resource predictions, reliability depends on the set of selected resources, not the performance of any single one. Second, resource assignment now requires identifying the *top-*$D$ resources for the user rather than performing a simple thresholding or single-winner selection. Third, effective gradient-based training demands a differentiable surrogate for the inherently discrete "at least $D$ resources are good" condition, and training objectives must capture aggregate reliability rather than per-resource accuracy.

In this paper, we address these challenges by proposing a complete end-to-end ML-assisted bulk allocation framework. In particular, we extend the outage-based learning framework from single-resource allocation to a bulk resource allocation

This work was supported by EPSRC NIA under Grant Reference Number UKRI917. *Corresponding author: Amir Masoud Molaei.*

Amir Masoud Molaei, Nidhi Simmons, Hien Quoc Ngo and Okan Yurduseven are with the Centre for Wireless Innovation, Queen's University Belfast, BT3 9DT Belfast, U.K. (e-mails: {a.molaei, nidhi.simmons, hien.ngo, okan.yurduseven}@qub.ac.uk).

David E. Simmons is with Dhali Holdings Ltd., BT5 7HW Belfast, U.K. (e-mail: dr.desimmons@gmail.com).

setting, where the user requires multiple resources simultaneously to meet a target rate. Unlike single-resource allocation, bulk allocation introduces a set-level reliability requirement, rendering conventional pointwise loss functions suboptimal. Our contributions are:

- We introduce a system model and formulate outage reliability tailored to ML-assisted bulk allocation.
- We design an operationally realistic and efficient allocation algorithm (decision rule), called gate + top-$D$ allocation (GTBA), based on ML-driven ranking.
- We develop a new differentiable loss function (ranking-aware bulk outage loss (RBOL)), trained using a differentiable surrogate of the GTBA outage event, which encourages the model to produce rankings ensuring that the selected resource set meets the required reliability level.
- We derive an exact reliability analysis for GTBA-based bulk allocation, decomposing the bulk outage probability (BOP) into gate and selection failure events, establishing an oracle lower bound, and characterizing asymptotic behavior that clarifies the fundamental performance limits and dominant failure mechanisms.

Together, these components close the gap between existing single-resource outage-based methods and the needs of modern multi-resource 6G systems, providing a practical and analytically grounded framework for ML-assisted bulk resource allocation.

The remainder of this paper is organized as follows. Section II introduces the system model and formalizes the definition of bulk outage. Section III presents the ML-assisted bulk allocation framework, including the GTBA policy and the proposed RBOL. In Section IV, we provide an analytical reliability analysis, establishing fundamental bounds and clarifying the roles of gate failures, selection failures and oracle performance. Section V reports extensive simulation results under various operating regimes, illustrating the benefits of RBOL relative to conventional loss functions. Finally, Section VI concludes the paper and outlines directions for future work.

## II. System Model and Bulk Outage Definition

We consider a wireless system in which a user has access to a collection of $R$ independent resources, indexed by $\mathcal{R}=\{1,2,\ldots,R\}$. Each resource $i$ corresponds to a channel realization $H_i(t,k)$ [8, 9], which determines a future mutual information or rate $C_i \triangleq C\left(H_i(t+l,l)\right)$ [8, 9], where $t$, $k$ and $l$ represent time, length of past channel samples, and length of future (unknown) samples, respectively, and $i \in \mathcal{R}$. The function $C(\cdot)$ captures the achievable rate [9] under a given modulation and coding scheme. As in outage-based systems, a resource is deemed "good" if the achieved rate exceeds a target threshold $\gamma_{\text{th}}$ (i.e., $C_i \geq \gamma_{\text{th}}$), and "bad" (in outage) otherwise (i.e., $C_i < \gamma_{\text{th}}$).

Fig. 1 presents a schematic of the system model. A bulk allocation request of size $D$ is said to be successful if at least $D$ allocated resources satisfy $C_i \geq \gamma_{\text{th}}$. Otherwise, a bulk outage occurs. In other words, a user requiring $D$ resources is considered to experience an outage if the total number of good resources in the allocated set is strictly less than $D$, i.e., $\#\{i \in A : C_i \geq \gamma_{\text{th}}\} < D$, where $A \subseteq \mathcal{R}$ denotes the set of assigned resources. The ML model produces, for each resource $i$, a predicted outage score (risk score) $q_i(\Theta) \triangleq Q\left(H_i(t,k);\Theta\right)$, where $Q(\cdot)$ and $\Theta$ represent the ML predictor and trainable (model) parameters, respectively, and $q_i(\Theta) \in [0,1]$. Lower values of $q_i(\Theta)$ indicate a higher predicted probability that the resource is good. Note that $q_i(\Theta)$ is a risk score calibrated to correlate with outage probability, not necessarily a perfectly calibrated probability.

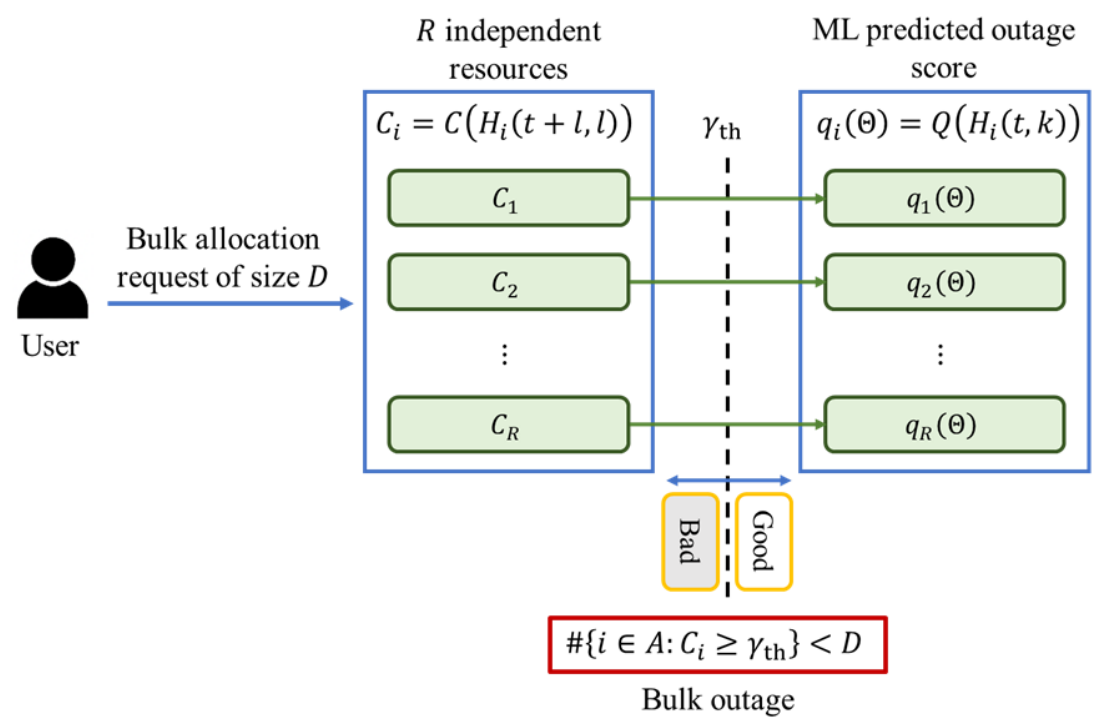

**Fig. 1.** Schematic of the system model and bulk outage definition.

This model structure generalizes prior single-resource frameworks to scenarios where users require multiple resources. However, because reliability now involves the aggregate behavior of selected resources, the training objective must reflect system-level performance rather than per-resource accuracy.

## III. ML-Assisted Bulk Allocation and Custom Loss

This section describes the proposed ML-assisted bulk allocation framework and the associated training objective. We first introduce a practical allocation policy, GTBA, which combines threshold-based admission control with ranking-based selection. We then develop a custom loss function that enables end-to-end training of ML models under this policy. Unlike conventional pointwise losses, the proposed formulation explicitly targets set-level reliability, ensuring that the selected group of resources jointly satisfies the bulk outage requirement.

### A. GTBA Policy

As mentioned in the previous sections, the bulk allocation model applies to scenarios in which multiple parallel resources must jointly meet a reliability target, such as carrier aggregation across subbands, multi-connectivity link selection for URLLC, multi-beam/antenna activation and multi-channel sensing. In these settings, the system must admit and rank candidates to reliably obtain at least $D$ usable resources, making decision-aligned learning practically important and meaningful. In this work, resource allocation is performed according to a simple but effective rule, which we refer to as the GTBA algorithm. It is a two-stage selection rule that combines threshold-based admission control with ranking-based bulk allocation. This structure reflects practical scheduling systems in which evidently unreliable resources are first excluded, and fine-grained ranking is then applied only among admissible candidates. In fact, we adopt GTBA because, firstly, it is practically implementable and secondly, it exposes analyzable failure modes (gate vs selection), enabling exact reliability characterization.

In the first stage, a fixed acceptance threshold (reference threshold) $q_{\text{th}} \in (0,1)$ is applied to all predicted risk scores. The gate threshold $q_{\text{th}}$ is a design parameter selected according to the desired reliability-feasibility tradeoff. The choice of $q_{\text{th}}$ depends indirectly on $\gamma_{\text{th}}$, since $\gamma_{\text{th}}$ determines the outage labels and therefore changes the distribution and calibration of the predicted risk scores $q_i(\Theta)$. Given the predicted risk scores $\{q_i(\Theta)\}_{i=1}^{R}$, the set of admissible resources is defined as

$$\mathcal{A}(q_{\text{th}}) \triangleq \{ i \in \mathcal{R} : q_i(\Theta) \le q_{\text{th}} \}. \quad (1)$$

Resources whose predicted risk exceeds $q_{\text{th}}$ are discarded and cannot be selected. This gate prevents the allocation of resources that the predictor deems highly unreliable, thereby improving robustness under adverse channel conditions. If the number of admissible resources is insufficient, i.e., $|\mathcal{A}(q_{\text{th}})| < D$, the system immediately declares a bulk outage. This event is referred to as a *gate failure* and corresponds to the inability of the predictor to identify at least $D$ potentially reliable resources.

If the gate condition $|\mathcal{A}(q_{\text{th}})| \ge D$ is satisfied, the system proceeds to the second stage. The admissible resources in $\mathcal{A}(q_{\text{th}})$ are ranked in ascending order of predicted risk (from most to least promising) as

$$\mathbf{s} \triangleq \underset{i \in \mathcal{A}(q_{\text{th}})}{\operatorname{argsort}} \left( q_i(\Theta) \right). \quad (2)$$

The final allocated set is formed by selecting the $D$ lowest-risk admissible resources as

$$\mathcal{A}_D(\Theta) \triangleq \{ \mathbf{s}(1), \mathbf{s}(2), \ldots, \mathbf{s}(D) \}, \quad (3)$$

where $\mathbf{s}(1)$ is the resource with the smallest predicted risk. Note that the top-$D$ selection is performed exclusively within the admissible set $\mathcal{A}(q_{\text{th}})$.

This ranking-based selection maximizes the predicted reliability of the allocated set while respecting the admission constraint imposed by the gate. This mechanism is attractive because of its simplicity, interpretability, and compatibility with ML models producing independent per-resource predictions. It also mirrors classical scheduling rules that select the best channels. However, unlike pure thresholding-based approaches, the GTBA algorithm exploits fine-grained risk ordering when many resources are admissible. Also, unlike pure top-$D$ selection, it prevents highly risky resources from being selected when all predictions are poor.

*B. Gate Failure and Bulk Outage Probabilities Under GTBA*

Formally, we define a binary indicator

$$g_i \triangleq [\![ C_i \ge \gamma_{\text{th}} ]\!] = 1 - y_i, \quad (4)$$

so that $g_i = 1$ represents a reliable resource, $[\![\cdot]\!]$ denotes Iverson bracket, and

$$y_i = \begin{cases} 1, & C_i < \gamma_{\text{th}}, \\ 0, & \text{otherwise}, \end{cases} \quad (5)$$

denotes the true outage indicator of resource $i$.

A gate failure under the GTBA algorithm occurs if fewer than $D$ resources pass the gate, i.e.,

$$E_{\text{gate}} = \{ |\mathcal{A}(q_{\text{th}})| < D \}. \quad (6)$$

This event captures failures due to overly pessimistic predictions or overly strict thresholding, in which the system cannot even form a candidate set of size $D$. Thus, the gate failure probability (GFP) is defined as

$$P_{\text{gate}}(D) = \Pr\left( |\mathcal{A}(q_{\text{th}})| < D \right). \quad (7)$$

If the gate condition $|\mathcal{A}(q_{\text{th}})| \ge D$ is satisfied, the system selects $\mathcal{A}_D(\Theta)$ according to (3). A selection failure occurs if fewer than $D$ of the selected resources are physically reliable, i.e.,

$$E_{\text{sel}} = \left\{ \sum_{i \in \mathcal{A}_D(\Theta)} g_i < D \right\}, \quad \mathcal{A}_D(\Theta) \subseteq \mathcal{A}(q_{\text{th}}), \quad |\mathcal{A}_D(\Theta)| = D. \quad (8)$$

This event reflects ranking errors within the admissible set, whereby unreliable resources are ranked ahead of reliable ones near the selection boundary.

A bulk outage under GTBA occurs if either a gate failure or a selection failure takes place. The bulk outage event is therefore defined as

$$E_{\text{bulk}} = E_{\text{gate}} \cup \left( E_{\text{sel}} \cap \{ |\mathcal{A}(q_{\text{th}})| \ge D \} \right). \quad (9)$$

The corresponding BOP is

$$P_{\text{bulk}}(D) = \Pr\left( |\mathcal{A}(q_{\text{th}})| < D \cup \sum_{i \in \mathcal{A}_D(\Theta)} g_i < D \right). \quad (10)$$

By construction, the gate failure event is a subset of the bulk outage event, which implies $P_{\text{gate}}(D) \le P_{\text{bulk}}(D)$. Moreover, the BOP can be decomposed as

$$P_{\text{bulk}}(D) = P_{\text{gate}}(D) + \Pr\left( |\mathcal{A}(q_{\text{th}})| \ge D, \sum_{i \in \mathcal{A}_D(\Theta)} g_i < D \right), \quad (11)$$

where the second term represents the probability of selection failure conditioned on passing the gate.

*C. Learning Objective and Loss Function*

In this subsection, we propose an RBOL, explicitly aligned with the GTBA reliability objective, which combines three components:
1. Soft gate and shortfall term (gate-failure control): ensuring at least $D$ reliable resources are softly selected.
2. Cutoff-aware ranking term (selection-failure control): discouraging bad resources from entering the top-$D$ set while good resources remain outside.
3. Per-resource BCE regularizer (stability term): to stabilize training for small $D$.
The details of these components are described below.

Training ML models under the GTBA algorithm is challenging because bulk outage depends on discrete operations: sorting, selecting the top-$D$ elements, and counting the number of reliable ones. These operations are non-differentiable [19] and cannot be directly embedded into a gradient-based learning framework. To overcome this, we introduce a differentiable surrogate of the bulk-outage condition.

We begin by defining a soft acceptance probability

$$p_i(\Theta) \triangleq \sigma\left(\left(q_{\text{th}} - q_i(\Theta)/\tau\right)\right), \quad \tau > 0, \tag{12}$$

where $\sigma(x) = 1/\left(1+e^{-x}\right)$ is the sigmoid function, and $\tau$ controls the sharpness of the transition. For small $\tau$, $p_i(\Theta)$ closely approximates the hard indicator $\left[\!\left[ q_i(\Theta) \le q_{\text{th}} \right]\!\right]$, so the soft gate closely matches the hard GTBA gate. This allows us to compute a differentiable surrogate of the number of reliable resources admitted by the hard GTBA gate. When the predicted risk score $q_i(\Theta)$ is much smaller than the threshold $q_{\text{th}}$ (indicating that the resource appears very safe), the soft selection probability satisfies $p_i(\Theta) \approx 1$. In this case, the model strongly favors selecting that resource. Conversely, when the predicted risk score $q_i$ is much larger than the threshold (indicating that the resource appears risky), the soft selection probability becomes $p_i(\Theta) \approx 0$, meaning the resource is unlikely to be selected.

If a resource is physically good, then $g_i = 1$, and the product $p_i g_i$ becomes a soft indicator that resource $i$ is both predicted to be good and is actually reliable. The soft (differentiable) estimate of the number of accepted and reliable resources is approximated by

$$\mathcal{G}(\Theta) = \sum_{i=1}^{R} p_i(\Theta)(1 - y_i). \tag{13}$$

In fact, $\mathcal{G}(\Theta)$ provides a soft (differentiable) estimate of the number of good resources that the model would select under the hard gate. Gate failures correspond to the event $\mathcal{G}(\Theta) < D.$ In other words, to satisfy the system requirement, we desire $\mathcal{G}(\Theta) \ge D$. The degree to which this requirement is violated is captured by the bulk shortfall term $\Delta_D \triangleq D - \mathcal{G}(\Theta)$. If $\Delta_D > 0$, we do not have enough good resources; and if $\Delta \le 0$, requirement is satisfied. So, a natural loss is $\phi\left(D - \mathcal{G}(\Theta)\right)$, where $\phi(\cdot)$ is some smooth increasing penalty. Therefore, soft-threshold bulk outage loss penalizes positive shortfall values using the softplus function as

$$\mathcal{L}_{\text{shortfall}}(\Theta) = \operatorname{softplus}(\Delta_D) = \ln\left(1 + e^{D - \mathcal{G}(\Theta)}\right). \tag{14}$$

This function is fully differentiable, convex in $\Delta_D$, and closely approximates $\max\{0, \Delta_D\}$. It grows smoothly when fewer than $D$ reliable resources pass the gate and is small when the gate condition is satisfied.

*Remark 1:* The indicator $g_i$ (equivalently $y_i$) is used only during training to define the supervised objective in (13) and (14). At deployment, neither $g_i$ nor $y_i$ is available or used; GTBA operates solely on the predicted scores $\{q_i\}$ and the fixed threshold $q_{\text{th}}$. Therefore, this does not introduce target leakage, i.e., the labels are not part of the model input nor the test-time decision rule, and they only shape $\Theta$ through offline optimization.

While the shortfall loss encourages selecting a sufficient number of reliable resources, it does not explicitly enforce correct ordering near the top-$D$ boundary, which is critical for GTBA. To address this, the second component of RBOL is needed. Let the predicted risks be sorted in ascending order $\mathbf{s}(1) \le \mathbf{s}(2) \le \ldots \le \mathbf{s}(D) \le \mathbf{s}^{\text{c}}(1) \le \mathbf{s}^{\text{c}}(2) \le \ldots \le \mathbf{s}^{\text{c}}(R-D)$, where

$$\mathbf{s}^{\text{c}} \triangleq \underset{i \in \mathcal{A}^{\text{c}}(q_{\text{th}})}{\operatorname{argsort}}\left(q_i(\Theta)\right), \tag{15}$$

$$\mathcal{A}^{\text{c}}(q_{\text{th}}) \triangleq \left\{ i \in \mathcal{R} : q_i(\Theta) > q_{\text{th}} \right\}, \tag{16}$$

and let

$$\mathcal{A}_D^{\text{c}}(\Theta) \triangleq \left\{ \mathbf{s}^{\text{c}}(1), \mathbf{s}^{\text{c}}(2), \ldots, \mathbf{s}^{\text{c}}(R-D) \right\}, \tag{17}$$

i.e., $\mathcal{A}_D^{\text{c}}(\Theta) = \mathcal{A}(q_{\text{th}}) \setminus \mathcal{A}_D(\Theta)$. Define

$$\begin{aligned} q_{\max}^{\text{sel}}(\Theta) &\triangleq \max_{i \in \mathcal{A}_D(\Theta)} q_i(\Theta), \\ q_{\min}^{\text{unsel}}(\Theta) &\triangleq \min_{i \in \mathcal{A}_D^{\text{c}}(\Theta)} q_i(\Theta), \end{aligned} \tag{18}$$

where $q_{\max}^{\text{sel}}(\Theta)$ is the largest risk among selected resources, and $q_{\min}^{\text{unsel}}(\Theta)$ is the smallest risk among unselected resources. Selection failures arise when unreliable resources appear inside $\mathcal{A}_D(\Theta)$ or reliable ones remain outside. For correct ranking with a cutoff margin $m > 0$, we want

$$q_{\max}^{\text{sel}}(\Theta) + m \le q_{\min}^{\text{unsel}}(\Theta), \tag{19}$$

i.e., all selected risks should be *safely below* the best unselected one. That means the violation quantity is $q_{\max}^{\text{sel}}(\Theta)+m-q_{\min}^{\text{unsel}}(\Theta)$, which is positive when things are bad (margin violated) and non-positive when things are good (margin satisfied). So, the ranking-aware cutoff penalty (loss) is defined as

$$\mathcal{L}_{\text{cut}}(\Theta)=\omega(\Theta)\,\text{softplus}\left(\left(q_{\max}^{\text{sel}}(\Theta)+m\right)-q_{\min}^{\text{unsel}}(\Theta)\right), \quad (20)$$

so that it is near zero when the inequality holds. In (20), $\omega(\Theta)$ is a smooth weight proportional to the product of the fraction of good resources outside the selected set $\mathcal{A}_D(\Theta)$ and the fraction of bad resources inside $\mathcal{A}_D(\Theta)$, i.e.,

$$\omega(\Theta)=\left(\frac{1}{\left|\mathcal{A}_D^{c}(\Theta)\right|}\sum_{i\in\mathcal{A}_D^{c}(\Theta)} g_i\right)\left(1-\frac{1}{\left|\mathcal{A}_D(\Theta)\right|}\sum_{i\in\mathcal{A}_D(\Theta)} g_i\right). \quad (21)$$

The cutoff ranking term $\mathcal{L}_{\text{cut}}(\Theta)$ directly penalizes ranking errors at the top-$D$ boundary and targets selection failures, which cause bulk outage under GTBA.

In addition to the set-level terms, we include a small BCE regularizer term

$$\mathcal{L}_{\text{BCE}}(\Theta)=\frac{1}{R}\sum_{i=1}^{R}\left[-y_i\ln q_i(\Theta)-(1-y_i)\ln\left(1-q_i(\Theta)\right)\right]. \quad (22)$$

This term is auxiliary and is not part of the GTBA objective. It stabilizes training, particularly for small values of $D$, and prevents degenerate score collapse. Its weight is adaptive, being larger for small $D$ and smaller for large $D$. Note that the label is Bernoulli (outage versus no outage) and the model output is interpreted as a probability-like score in $[0,1]$. BCE is the standard proper scoring rule for probabilistic binary prediction [20, 21] and provides strong, well-shaped gradients, especially when predictions saturate near 0 or 1, exactly when the bulk terms can become noisy or flat (particularly for small $D$).

Finally, the proposed RBOL is defined as

$$\begin{aligned}&\mathcal{L}_{\text{RBOL}}(\Theta)=\\&\quad\mathcal{L}_{\text{shortfall}}(\Theta)+\lambda_{\text{rank}}\mathcal{L}_{\text{cut}}(\Theta)+\lambda_{\text{bce}}(D)\mathcal{L}_{\text{BCE}}\left(g_i,q_i(\Theta)\right),\end{aligned} \quad (23)$$

where ranking weight $\lambda_{\text{rank}}>0$ controls the emphasis on ranking correctness, and BCE weight $\lambda_{\text{bce}}(D)>0$ is a small stabilizing weight (adaptive in $D$). Then the training objective (expected risk) is

$$\min_{\Theta}\mathbb{E}\left[\mathcal{L}_{\text{RBOL}}(\Theta)\right], \quad (24)$$

where the expectation $\mathbb{E}[\cdot]$ is taken over the randomness of the training data (channel realizations and the generator [9]). In fact, the expectation in (24) represents the population risk over channel realizations; in practice, it is minimized via stochastic gradient descent using Monte-Carlo samples produced by the data generator.

RBOL therefore aligns training directly with the system's objective of ensuring that the top-$D$ resources selected by the GTBA algorithm are reliable. Unlike single-resource outage losses, which treat each resource independently, RBOL considers the reliability of the selected resource set.

## IV. Analytical Reliability Analyses

In this section, we analyze the reliability performance of the proposed GTBA policy and establish fundamental bounds that clarify the roles of gate failures, selection failures and oracle performance. The analysis is intentionally algorithm-centric and aligns directly with the operational definition of bulk outage introduced in Section III.

### *A. Decomposition of BOP*

Recall that under GTBA, a bulk outage occurs if either the gate admits fewer than $D$ resources or, conditional on passing the gate, the selected set contains fewer than $D$ physically reliable resources. Using the definitions in Section III-B and the law of total probability [22], the BOP can be written as

$$\begin{aligned}P_{\text{bulk}}(D)=&\Pr\left(\left|\mathcal{A}(q_{\text{th}})\right|<D\right)\\&+\Pr\left(\sum_{i\in\mathcal{A}_D(\Theta)} g_i<D\,\middle|\,\left|\mathcal{A}(q_{\text{th}})\right|\geq D\right)\Pr\left(\left|\mathcal{A}(q_{\text{th}})\right|\geq D\right),\end{aligned} \quad (25)$$

where, according to (1),

$$\left|\mathcal{A}(q_{\text{th}})\right|=\sum_{i=1}^{R}\left[\!\left[q_i(\Theta)\leq q_{\text{th}}\right]\!\right], \quad (26)$$

is the number of resources admitted by the gate (number of accepted resources (NAR)) for a given realization. The first term in (26) corresponds to the GFP and captures errors arising from overly pessimistic predictions or an excessively strict threshold $q_{\text{th}}$. The second term captures selection failures, which occur when ranking errors cause unreliable resources to enter the top-$D$ set despite sufficient admissible candidates. This decomposition is exact and mirrors the operational logic implemented in the evaluation algorithm: gate failures are counted whenever fewer than $D$ resources satisfy $q_i(\Theta)\leq q_{\text{th}}$, while selection failures are counted only when the gate succeeds, but fewer than $D$ selected resources are physically reliable.

The GFP depends only on the marginal distribution of the predicted risk scores $\{q_i(\Theta)\}$ and the threshold $q_{\text{th}}$, and not on the internal ranking among admissible resources. For a fixed predictor, the GFP is monotonically non-decreasing in $D$, since requiring more admissible resources makes the gate condition harder to satisfy. Importantly, training with RBOL directly suppresses gate failures through the shortfall term in (14), which encourages the expected number of accepted and reliable resources to exceed $D$. This explains why, in the numerical results in Section V-C, the RBOL-trained model exhibits significantly lower GFPs compared to pointwise-trained models, even when evaluated under the same GTBA rule.

Conditioned on successful gating, a selection failure occurs when the top-$D$ ranked admissible resources do not contain $D$ reliable ones. This event depends on the relative ordering of predicted risk scores across good and bad resources, rather than on their absolute values. According to (18) and (19), selection failures are strongly associated with events where $q_{\min}^{\text{unsel}}(\Theta) < q_{\max}^{\text{sel}}(\Theta)$, or $q_{\max}^{\text{sel}}(\Theta) + m > q_{\min}^{\text{unsel}}(\Theta)$ when using a margin, particularly when the unselected resource is physically good and the selected resource is physically bad. The ranking-aware cutoff term in RBOL explicitly penalizes ranking boundary violations at the top-$D$ cutoff, which are known to contribute to bulk outage under GTBA.

*B. Oracle BOP (OBOP): Lower Bound*

To benchmark the best achievable reliability under the same physical channel realizations, we define an oracle allocator that has perfect knowledge of the true outage indicators $\{g_i\}_{i=1}^{R}$. The oracle selects the $D$ best physically reliable resources whenever possible.

A bulk outage under the oracle occurs if and only if the total number of physically good resources is smaller than $D$, i.e.,

$$E_{\text{oracle}}(D) = \left\{ \sum_{i=1}^{R} g_i < D \right\}. \tag{27}$$

Hence, the OBOP is

$$P_{\text{oracle}}(D) = \Pr\left( \sum_{i=1}^{R} g_i < D \right). \tag{28}$$

Importantly, the oracle never suffers from gate or selection failures; it fails only when the physical channel conditions themselves are insufficient. This quantity is therefore independent of any ML prediction errors and depends solely on the underlying channel statistics.

*C. Oracle as a Fundamental Lower Bound*

The oracle outage probability provides a strict lower bound on the performance of any practical GTBA-based scheme.

*Lemma 1 (Oracle Lower Bound):* For any choice of predictor $q_i(\Theta)$, threshold $q_{\text{th}}$, and ranking rule, the bulk outage probability under GTBA satisfies

$$P_{\text{bulk}}(D) \geq P_{\text{oracle}}(D). \tag{29}$$

*Proof:* If $\sum_{i=1}^{R} g_i < D$, no allocation (ML-based or otherwise) can select $D$ reliable resources. Thus, every scheme must declare a bulk outage whenever the oracle does. Conversely, GTBA may incur additional outages due to gate or ranking errors even when $\sum_{i=1}^{R} g_i \geq D$. Hence, the oracle outage event is a subset of the GTBA outage event, which proves the inequality. ■

Note that for any ML-assisted allocation policy (including GTBA), the selected set $\mathcal{A}_D(\Theta)$ is a subset of the available resources. Therefore, $\sum_{i \in \mathcal{A}_D(\Theta)} g_i \leq \sum_{i=1}^{R} g_i$, which immediately implies (29).

Thus, the oracle curve constitutes a strict lower bound on achievable bulk outage probability. Any gap between $P_{\text{bulk}}(D)$ and $P_{\text{oracle}}(D)$ quantifies the combined impact of imperfect prediction, thresholding and ranking. This bound explains why the oracle curve in Section V-C consistently lies below all learned methods and why its gap relative to learned curves quantifies the aggregate impact of prediction and ranking errors.

*D. Asymptotic Behavior (Large-$R$, Fixed-$D$)*

For further analytical consideration, in this subsection, we consider the regime where the number of candidate resources $R$ grows while the bulk requirement $D$ remains fixed.

*Proposition 1 (Asymptotic Reliability Regime):* Assume the resources are independent and identically distributed, with

$$p_g \triangleq \Pr(g_i = 1) > 0, \tag{30}$$

where $p_g$ is the marginal probability that a randomly chosen resource is physically reliable. Then, as $R \to \infty$ with fixed $D$, $P_{\text{oracle}}(D) \to 0$.

*Proof:* By the law of large numbers [23],

$$\frac{1}{R} \sum_{i=1}^{R} g_i \xrightarrow{\text{almost surely}} p_g. \tag{31}$$

Hence, according to (28), the probability that fewer than $D$ successes occur among $R$ trials decays exponentially in $R$. ■

Note that, for GTBA-based policies, a sufficiently well-trained model with calibrated scores will also satisfy $P_{\text{gate}}(D) \to 0$, since the expected number of admissible resources grows linearly with $R$.

Therefore, in the large-$R$, fixed-$D$ regimes, BOP is dominated by ranking errors near the top-$D$ cutoff. This result implies that in rich-resource regimes, bulk outage is dominated not by physical scarcity but by algorithmic imperfections. This observation provides further justification for incorporating a ranking-aware penalty, as done in RBOL, rather than relying solely on pointwise classification accuracy.

*E. Reliability Implications of Gate Design*

The gate threshold $q_{\text{th}}$ plays a dual role. A smaller $q_{\text{th}}$ reduces the likelihood of admitting unreliable resources, thereby improving ranking purity within $\mathcal{A}(q_{\text{th}})$, but increases the probability of gate failure. Conversely, a larger $q_{\text{th}}$ reduces gate failures but increases the risk that unreliable resources enter the admissible set, raising selection failures.

The proposed RBOL directly targets this trade-off. The shortfall term in (14) penalizes violations of $|\mathcal{A}(q_{\text{th}})| \geq D$, thereby reducing $P_{\text{gate}}(D)$, while the cutoff-aware ranking term in (20) penalizes mis-ordering at the top-$D$ boundary, thereby reducing selection failures. This alignment explains why, in the numerical results in Section V-C, the RBOL-trained model exhibits substantially lower GFP than the other-trained models for the same $q_{\text{th}}$.

*F. Implications for Loss Design and Training*

The reliability analysis highlights three key insights:

1. Gate failures dominate conservative predictors, especially for small $R$ or poorly calibrated thresholds.
2. Selection failures become a dominant mechanism once gate feasibility is ensured.

3. RBOL explicitly targets both failure modes, whereas pointwise losses primarily reduce per-resource misclassification without directly controlling top-$D$ reliability.

As a result, minimizing RBOL aligns training with reducing the excess outage probability $P_{\text{bulk}}(D)-P_{\text{oracle}}(D)$, which represents the irreducible performance gap attributable to learning and ranking errors.

This analytical perspective explains the trends observed in Section V-C, where RBOL consistently approaches the oracle curve more closely than baseline losses across a wide range of $D$.

### G. Reliability Interpretation (Summary)

The reliability analysis demonstrates that (i) bulk outage under GTBA naturally decomposes into gate and selection failures, (ii) the oracle outage probability provides a strict lower bound on achievable performance, and (iii) for large candidate pools, ranking errors, not resource scarcity, become the dominant reliability bottleneck. The proposed RBOL is explicitly designed to address these mechanisms, thereby aligning the training objective with the system-level reliability metric of interest. The insights gained in this section motivate the numerical investigations presented in Section V.

## V. Simulation Results and Discussion

In this section, we evaluate the proposed ML-assisted bulk allocation framework and the RBOL through extensive Monte Carlo simulations [24]. Our objective is to compare the bulk outage reliability achieved by models trained with RBOL against conventional baselines, including mean absolute error (MAE) [9, 25], mean squared error (MSE) [9, 26] and BCE [9, 27], and also OLF [9], under a realistic and controlled channel evolution model. All results are obtained using the GTBA rule described in Section III.

### A. Data Generation and Channel Model

Training, validation and testing data are generated synthetically using the custom data generator implemented by the code developed in [28]. Each training sample corresponds to a single user with access to $R$ parallel and statistically independent resources. In all experiments, we fix $R=16$, which represents the number of candidate resources available to the user at a given scheduling instant. It is worth clarifying that although practical orthogonal frequency division multiplexing (new radio) systems [29] may expose hundreds of resource block (RBs)/subcarriers, real schedulers typically operate on a reduced candidate pool to meet real-time constraints; in long term evolution, for example, scheduling is commonly implemented in two phases, where a time-domain step first forms a small scheduling candidate set [30] and the frequency-domain scheduler then allocates RBs only within this set to reduce complexity. Accordingly, we interpret $R=16$ as the size of an effective candidate pool after such coarse filtering/preselection (e.g., based on continuous quality improvement (measurements)), rather than the total number of physical RBs in the bandwidth.

Each resource is modeled as a frequency-domain channel derived from a tapped-delay-line representation [9, 31] with $v=32$ complex Gaussian taps. The taps are drawn independently according to a circularly symmetric complex Gaussian distribution and normalized such that the total average channel power is unity. A fast Fourier transform is applied to obtain the frequency response, and equally spaced frequency samples are extracted to represent distinct resources. Temporal channel correlation [32] is introduced by applying small random phase rotations across successive samples, controlled by a phase-shift parameter set to 0.1 radians [9], which emulates moderate user mobility.

For each resource $i$, the ML model observes a sequence of $k=100$ past channel magnitude samples and predicts the likelihood of outage over the next $l=10$ future samples. The instantaneous achievable rate is computed using the standard Shannon expression [33] (for more details, see (2) from [11]). Unless otherwise specified, the signal-to-noise ratio (SNR) is fixed at 0 dB. Also, a resource is declared reliable if its future rate exceeds the target threshold $\gamma_{\text{th}}=1.2$; otherwise, it is labeled as being in outage. These binary outage labels form the supervision signal used during training.

To ensure statistical robustness, training and validation data are generated on the fly [34] using statistically independent channel realizations across epochs, with the same realization sequence used for all models within each experiment, ensuring fair comparison across loss functions. Each training epoch consists of 60 batches, each corresponding to one realization of the $R$-resource system. Over 65 epochs, each model is therefore trained on 3,900 statistically independent realizations per experiment, with an equal number used for validation. Performance evaluation is carried out on a separate test set comprising 3,000 independent realizations. Each realization contains $R$ correlated resources, so the effective number of per-resource samples scales linearly with $R$. The predictor is retrained 10 times, and the average performance is used as a single data point for analysis.

### B. Model Architecture and Training Setup

All predictors share the same neural network (NN) architecture to ensure a fair comparison between loss functions. Specifically, each model consists of a single long short-term memory (LSTM) [35] layer with 16 hidden units, followed by a fully connected layer with 10 units and PReLU activation [36], and a final sigmoid output layer that produces a scalar risk score $q_i\in[0,1]$ for each resource. Lower values of $q_i$ indicate a lower predicted outage risk.

Models are trained using the Adam optimizer [37] with default TensorFlow/Keras parameters (i.e., learning rate $10^{-3}$, $\beta_1=0.9$, $\beta_2=0.999$ and $\varepsilon=10^{-7}$, where $\beta_1$, $\beta_2$ and $\varepsilon$ are the exponential decay rate for the first and second moments of the gradients, and a small positive constant added for numerical stability, respectively). We consider five training objectives: MAE, MSE and BCE (three conventional per-resource classification losses that treat each resource independently, OLF (a reliability-aware baseline derived from prior single-resource outage analysis [9, 10], and RBOL (the proposed loss function tailored to the GTBA policy and bulk reliability objective). For the bulk model, a separate network is trained for each required bulk size $D \in \{2,4,6,8,10\}$. Therefore, RBOL incurs additional offline training cost due to set-level operations (gate surrogates and cutoff terms). The acceptance threshold in the GTBA gate is fixed to $q_{\text{th}} = 0.4$ across all models and experiments. RBOL hyperparameters were tuned on a validation set using a small grid search and then kept fixed for all experiments. A $D$-dependent $\tau$ is a practical way to keep the soft surrogate neither too smooth nor too sharp across all $D$, leading to stable training and better alignment with the hard GTBA decision. So, in implementations, for small $D$, since the decision boundary is sparse and unstable, we use a larger $\tau$ (smoother) to avoid noisy gradients. However, for large $D$, since many items cluster near the cutoff, we need a smaller $\tau$ (sharper) to correctly separate the top-$D$ set.

A summary of the model architecture and all parameters and their values are given in Table I. Throughout Section V-C, unless otherwise noted, parameter values are as given in this table.

After training, all models are evaluated on the same shared test set using the hard GTBA decision rule. For each value of $D$, we measure the empirical BOP and GFP, as defined in Section III-B. We also compute the OBOP, which corresponds to perfect knowledge of future outages and serves as a fundamental lower bound. The following subsection presents and discusses the resulting reliability curves.

*C. Results*

In this subsection, we evaluate the reliability performance of the proposed RBOL under the GTBA policy in various experiments and regimes, and compare it with conventional training objectives. In addition to outage curves, we demonstrate that the proposed loss is effective, stable and aligned with the analytical reliability framework presented in Section IV.

*Example 1 (Bulk Reliability versus Required Set Size):* In the first experiment, the GFP and BOP are evaluated as functions of the required number of resources $D$ under the balanced-stress regime. Under this regime, $\gamma_{\text{th}}$ is moderate (assumed to be equal to 1.2); the system operates near a critical point where gate and ranking decisions matter. Fig. 2(a) depicts the probability of gate failure, defined as the event that fewer than $D$ resources satisfy $q_i \le q_{\text{th}}$. Several observations can be made. First, for the BCE- and finite-outage-trained models, the GFP increases rapidly with $D$, approaching unity already for moderate values of $D$. This behavior reflects the tendency of pointwise loss functions to produce conservative risk estimates, which causes the gate to reject an excessive number of resources when multiple reliable allocations are required. Similar behavior is observed for MAE- and MSE-trained models, whose gate failure probabilities closely track those of BCE and OLF across all values of $D$. In contrast, the proposed RBOL achieves a substantially lower GFP over the entire range of $D$. For example, at $D = 4$, RBOL reduces the GFP from approximately 0.29-0.37 (for BCE, MSE, MAE and OLF) to about 0.034, and at $D = 6$, from above 0.76 to approximately 0.14. This confirms that the shortfall component of RBOL effectively encourages the admission of at least $D$ potentially reliable resources, thereby suppressing gate failures. For very large $D$ (e.g., $D = 10$), all methods experience high GFP, reflecting the intrinsic difficulty of identifying many reliable resources in the balanced-stress regime. Nevertheless, RBOL significantly delays the onset of near-certain gate failure compared to all baseline losses.

TABLE I
MODEL ARCHITECTURE AND HYPERPARAMETERS FOR TRAINING AND CUSTOM LOSS

| Component | Parameter | Value / Description |
|---|---|---|
| **Input data** | $R$ | 16 |
| | $k$ | 100 samples |
| | $l$ | 10 samples |
| **Channel model** | $v$ | 32 complex Gaussian taps |
| | Channel normalization | Unit average power |
| | Temporal correlation | Phase rotation with step size 0.1 rad |
| **Rate model** | Achievable rate | $C$ (a function of $H$ and SNR [11]) |
| | SNR | 0 dB |
| | $\gamma_{\text{th}}$ | $\{1,1.2,1.4\}$ |
| **NN** | Architecture | LSTM → Dense (PReLU) → Dense (Sigmoid) |
| | LSTM units | 16 |
| | Dense layer (hidden) | 10 units, PReLU activation |
| | Output layer | 1 unit, Sigmoid activation |
| **Output** | $q_i$ | Outage risk score in $[0,1]$ |
| **Optimizer** | Type | Adam |
| | Learning rate | $10^{-3}$ |
| | $\beta_1$ | 0.9 |
| | $\beta_2$ | 0.999 |
| | $\varepsilon$ | $10^{-7}$ |
| **Training** | Epochs | 65 |
| | Batches per epoch | 60 |
| | Test samples | 3000 for each experiment |
| **GTBA policy** | $q_{\text{th}}$ | 0.4 |
| | $D$ | $\{2,4,6,8,10\}$ |
| **RBOL loss** | $\tau$ | 0.15 (if $D \le 2$), $\max(0.08, 0.2/D)$ (otherwise) |
| | $\lambda_{\text{rank}}$ | 8 |
| | $m$ | 0.08 |
| | $\lambda_{\text{bce}}$ | 0.2 (if $D \le 2$), 0.05 (otherwise) |

Fig. 2(b) reports the corresponding BOP under the GTBA decision rule. As expected from the analysis in Section IV, the BOP is always lower-bounded by the OBOP, which depends solely on channel statistics. For small values of $D$, all learning-based schemes perform close to the oracle. As $D$ increases, however, marked differences emerge. OLF, MAE, MSE and BCE losses exhibit a sharp increase in BOP, reaching outage probabilities close to one at moderate $D$.

This degradation is primarily driven by the rapid growth of gate failures, as evidenced in Fig. 2(a). Across all values of $D$, the proposed RBOL consistently achieves the lowest BOP among the learned approaches. At $D=4$, RBOL reduces BOP by roughly 27%-41% relative to the pointwise baselines, and at $D=6$, the reduction remains on the order of 15%-21%. Even for larger $D$, where outages become unavoidable for most schemes, RBOL remains noticeably closer to the oracle curve.

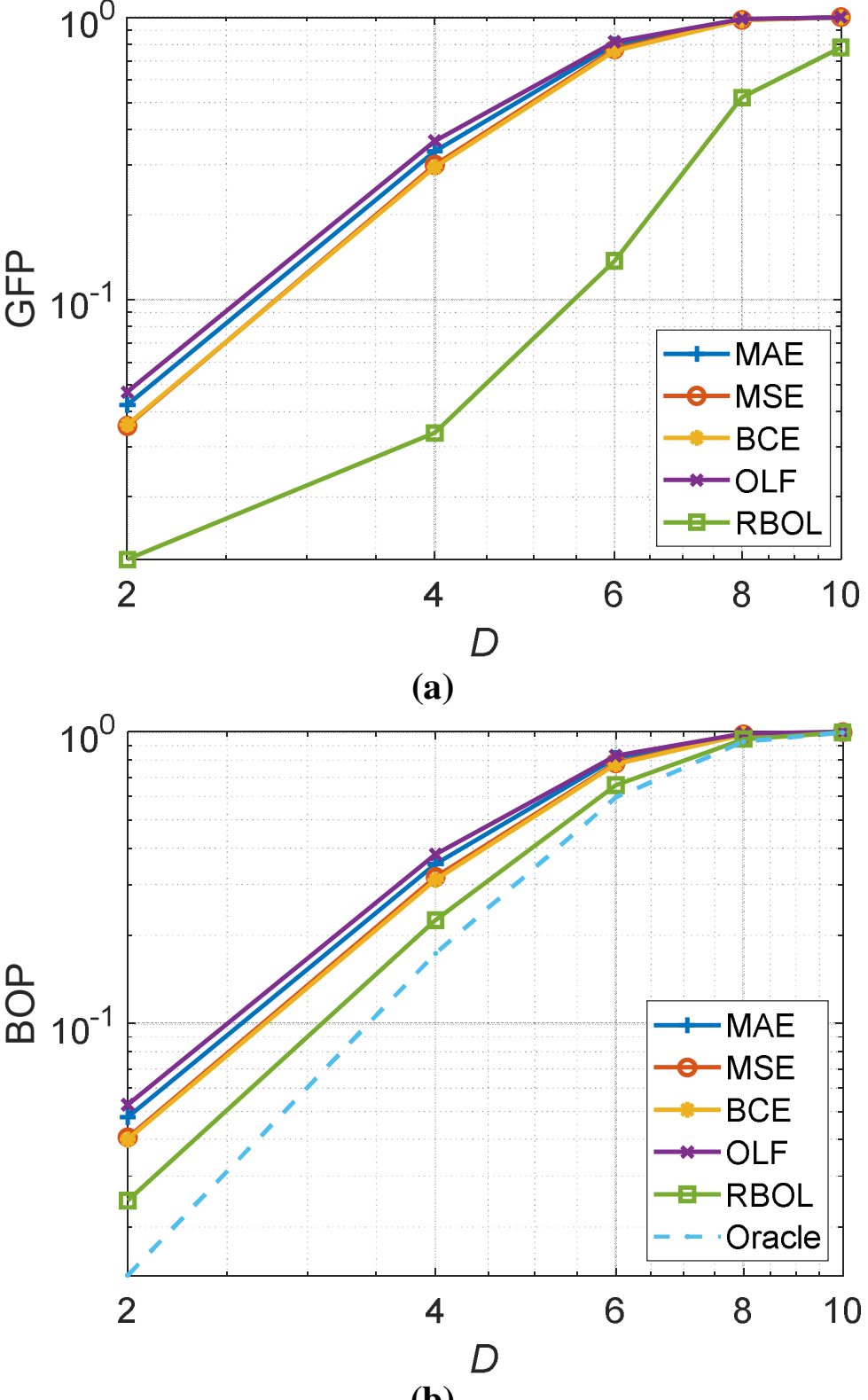

**Fig. 2.** Comparison of different loss functions versus $D$ changes in Example 1 when $\gamma_{\text{th}}=1.2$; **(a)** GFP; **(b)** BOP.

Note that the oracle curve in Fig. 2(b) represents the minimum achievable BOP dictated solely by the underlying channel statistics. The remaining gap between RBOL and the oracle, therefore, quantifies the irreducible impact of imperfect prediction and ranking. Importantly, RBOL consistently attains the smallest gap to the oracle across all values of $D$, confirming that its training objective is well aligned with the system-level reliability metric of interest. This observation is consistent with the analytical results of Section IV, which predict that gate and ranking errors dominate reliability loss in regimes where sufficient good resources exist. Overall, the results in Figs. 2(a) and 2(b) demonstrate that RBOL substantially suppresses gate failures and achieves lower bulk outage probabilities compared to conventional pointwise losses, including MAE, MSE, BCE and OLF. Losses that do not explicitly enforce set-level reliability fail to jointly control these mechanisms, leading to significantly degraded performance as the bulk requirement increases.

Finally, at the end of Example 1, to evaluate the robustness of the proposed RBOL in terms of hyperparameter sensitivity, we performed a small sweep over $\lambda_{\text{rank}} \in \{2,4,6,8\}$ and $m \in \{0.03, 0.05, 0.08\}$, keeping all other settings fixed. As shown in Table II, RBOL performance is stable across this range, with BOP varying only slightly (typically at the third decimal place) for moderate $D$, while consistently outperforming the BCE baseline. These results indicate that RBOL does not rely on fragile hyperparameter tuning and generalizes well over reasonable parameter choices.

TABLE II
RBOL HYPERPARAMETER SENSITIVITY IN EXAMPLE 1

| Loss function | $\lambda_{\text{rank}}$ | $m$ | BOP ($D=2$) | BOP ($D=4$) | BOP ($D=6$) | BOP ($D=8$) | BOP ($D=10$) |
|---|---|---|---|---|---|---|---|
| BCE | - | - | 0.04 | 0.312 | 0.776 | 0.981 | 1 |
| RBOL | 8 | 0.08 | 0.0247 | 0.224 | 0.654 | 0.943 | 0.998 |
| RBOL | 8 | 0.05 | 0.0247 | 0.224 | 0.655 | 0.944 | 0.998 |
| RBOL | 8 | 0.03 | 0.025 | 0.227 | 0.656 | 0.944 | 0.998 |
| RBOL | 6 | 0.08 | 0.0247 | 0.231 | 0.657 | 0.943 | 0.998 |
| RBOL | 4 | 0.08 | 0.025 | 0.228 | 0.655 | 0.944 | 0.998 |
| RBOL | 2 | 0.08 | 0.0253 | 0.226 | 0.66 | 0.943 | 0.998 |

*Example 2 (Impact of Stress Regime):* In this experiment, we examine the impact of the target rate threshold $\gamma_{\text{th}}$ on system reliability, thereby illustrating the behavior of the learning-based allocation schemes under different stress regimes. In addition to the balanced-stress case of Example 1 ($\gamma_{\text{th}}=1.2$), we consider a light-stress regime (reliable resources are abundant) with low $\gamma_{\text{th}}$ and a heavy-stress regime (reliable resources are scarce) with high $\gamma_{\text{th}}$. All other system parameters, network architectures and training settings are kept unchanged.

Fig. 3(a) depicts the GFP as a function of the required number of resources $D$ for the light-stress regime ($\gamma_{\text{th}}=1$). As expected, when channel conditions are favorable, all methods achieve low GFP for small $D$. For instance, at $D=2$, the GFP of all schemes is below 0.018. As $D$ increases, pointwise losses such as MAE, MSE, BCE and OLF exhibit a noticeable growth in GFP; at $D=6$, their GFP values range from approximately 0.14 to 0.65. In contrast, the proposed RBOL maintains a substantially lower GFP, reducing it to about 0.026 at $D=6$. Even at larger $D$, RBOL consistently admits more candidate resources than the baselines, delaying the onset of near-certain gate failure.

The corresponding BOP results for $\gamma_{\text{th}}=1$ are shown in Fig. 3(b). For small $D$, all learning-based methods perform close to the oracle bound. However, as $D$ increases, OLF exhibits a sharp rise in BOP. For example, at $D=8$, its BOP value approaches 0.95, whereas RBOL achieves a lower BOP of approximately 0.74, remaining closer to the oracle value of 0.68. By comparing all the graphs in Fig. 3(b), it is confirmed that even in light-stress conditions, explicitly accounting for the bulk requirement during training yields competitive and sometimes improved reliability (although with less severity) across the entire range of $D$.

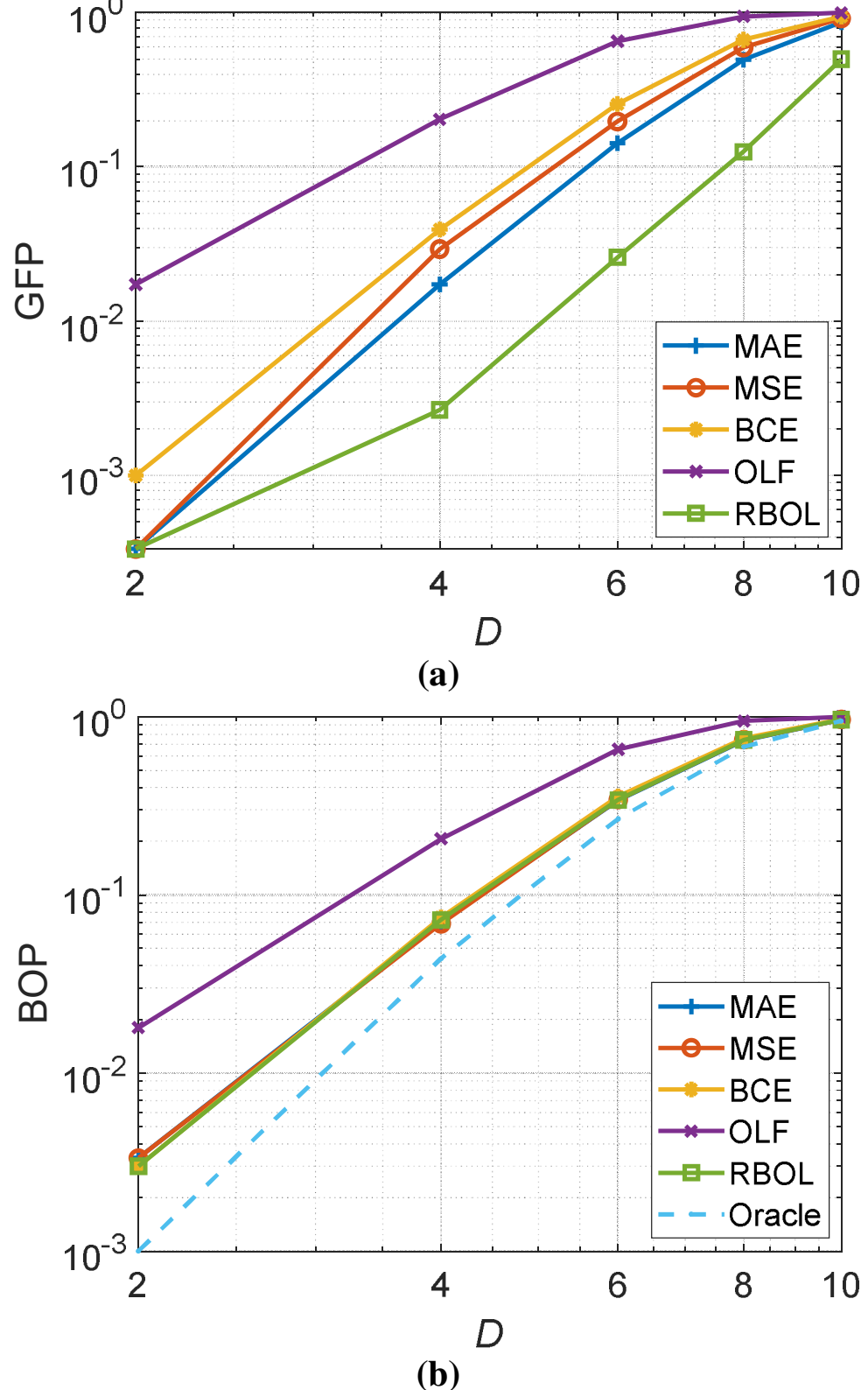

**Fig. 3.** Comparison of different loss functions versus $D$ changes in Example 2 when $\gamma_{\text{th}} = 1$; **(a)** GFP; **(b)** BOP.

Figs. 4(a) and 4(b) report the GFP and BOP, respectively, for the heavy-stress regime ($\gamma_{\text{th}} = 1.4$). In this case, the system operates close to its reliability limits. As seen in Fig. 4(a), GFP increases rapidly with $D$ for all methods. At $D = 4$, OLF, MSE, MAE and BCE already exhibit GFP values above 0.74. By contrast, RBOL limits the GFP to about 0.15 at $D = 4$ and 0.63 at $D = 6$, significantly postponing the onset of severe gate failures. The corresponding BOP curves in Fig. 4(b) show that although all methods approach outage probabilities close to one for moderate $D$, RBOL consistently achieves the lowest BOP across the entire range of $D$. For example, at $D = 4$, RBOL reduces the BOP to approximately 0.69, compared to values exceeding 0.77 for OLF, MSE, MAE and BCE. At $D = 6$, RBOL maintains a BOP of about 0.972, remaining closer to the oracle bound of 0.968 than the pointwise baselines.

Comparing Figs. 2 and 3 highlights a trend: as the stress level increases, the performance gap between RBOL and conventional pointwise losses becomes more pronounced. This behavior is consistent with the analytical insights of Section IV, which predict that set-level effects dominate reliability when the availability of good resources is limited. Loss functions that do not explicitly account for these effects fail to control gate behavior and ranking jointly, leading to rapid degradation in bulk reliability. However, considering all the results in Figs 2-4, it should be noted that at very high stress and/or large $D$, all curves saturate near 1, so there is less room for absolute improvements. That's why, numerically, RBOL's advantage is often largest in the mid-stress / mid-$D$ region. Overall, Examples 1 and 2 demonstrate that RBOL outperforms MAE/MSE/BCE/OLF across all stress regimes, with the largest absolute gains observed in the intermediate (balanced-stress) regime, while gains diminish in the heavy-stress regime due to outage saturation.

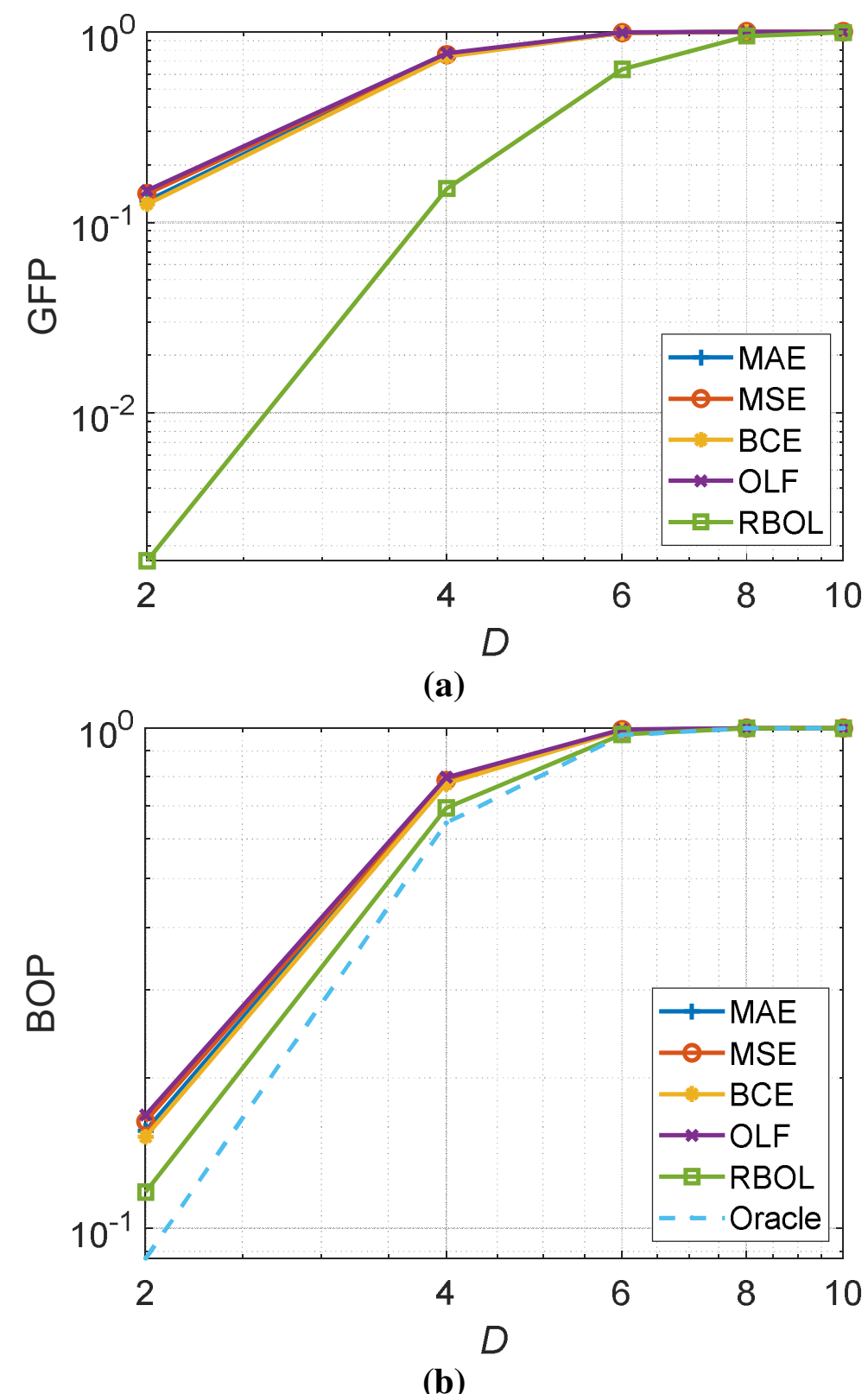

**Fig. 4.** Comparison of different loss functions versus $D$ changes in Example 2 when $\gamma_{\text{th}} = 1.4$; **(a)** GFP; **(b)** BOP.

*Example 3 (Robustness to SNR Variations):* In this experiment, we evaluate the robustness of the proposed learning-based allocation scheme to changes in channel quality. Specifically, we fix the target rate threshold at $\gamma_{\text{th}} = 1.2$ and examine the GFP and BOP as functions of the SNR. The models are trained once at the nominal operating point and then evaluated under an SNR sweep of $\{-6, -3, 0, 3, 6\}$ dB using a fixed gate threshold $q_{\text{th}} = 0.4$.

Fig. 5 shows the results for $D = 4$. At very low SNR $(-6 \text{ dB})$, all schemes experience near-certain outage and gate failure, with GFP and BOP equal to one. At $-3$ dB, RBOL already begins to outperform the pointwise baselines: while OLF, MAE, MSE and BCE exhibit BOP values above 0.99, RBOL reduces the BOP to approximately 0.96, remaining closer to the oracle value of 0.95. The corresponding GFP at $-3$ dB is about 0.71 for RBOL, compared to values exceeding 0.99 for the pointwise losses. At the nominal SNR of 0 dB, pronounced differences emerge. As shown in Fig. 5(a), RBOL achieves a BOP of approximately 0.23, compared to 0.31 for BCE, 0.36 for OLF, and similar values for MAE and MSE, representing a relative reduction of roughly 26%-36%. This gain is primarily driven by a strong suppression of gate failures: Fig. 5(b) indicates that RBOL reduces the GFP to about 0.037, whereas the pointwise losses exhibit GFP values around 0.29-0.34. At higher SNRs (3 dB and 6 dB), all methods rapidly approach the oracle bound and achieve negligible outage, indicating that the system is no longer reliability-limited in this regime.

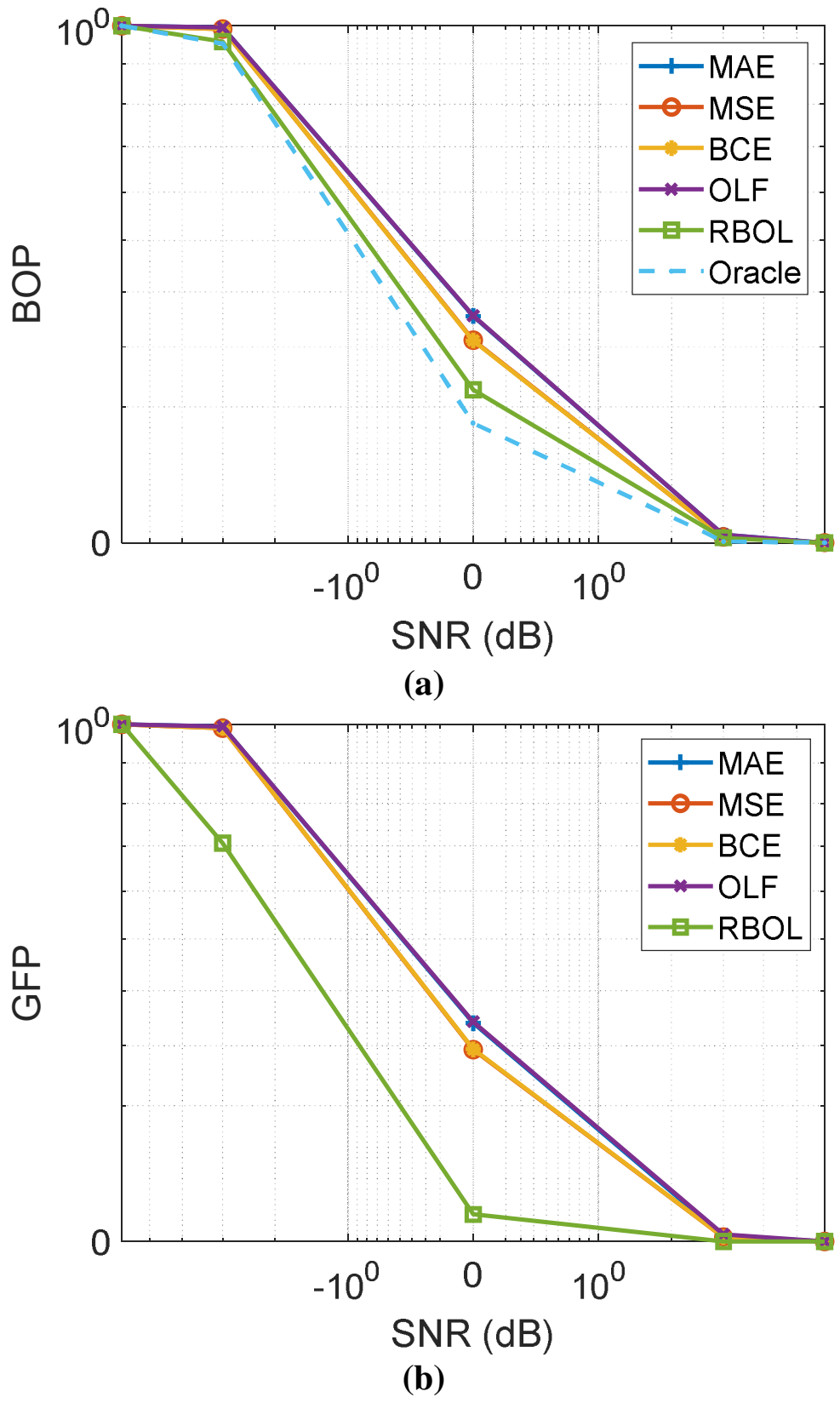

**Fig. 5.** Comparison of different loss functions versus SNR in Example 3 for $D=4$; **(a)** BOP, **(b)** GFP.

The corresponding results for $D=6$ are shown in Fig. 6. As expected, increasing the bulk requirement makes the problem more challenging. At −3 dB, all methods exhibit BOP values equal to one, reflecting the severe stress imposed by both low SNR and large $D$. At 0 dB, however, RBOL again provides a clear advantage. Fig. 6(a) shows that RBOL reduces the BOP to approximately 0.66, compared to 0.78 for BCE, 0.83 for OLF, and similar values for MSE and MAE, yielding an absolute improvement of about 0.12-0.17. This improvement is accompanied by a dramatic reduction in GFP: at 0 dB, RBOL achieves a GFP of roughly 0.2, whereas the pointwise losses suffer GFP values in the range 0.76-0.82 (see Fig. 6(b)). At 3 dB, RBOL continues to outperform the baselines, achieving a BOP of about 0.068, compared to 0.082-0.11 for other losses, and remaining closer to the oracle value of 0.032. At 6 dB, all methods achieve near-zero GFP and BOP, indicating that reliable operation is guaranteed when channel conditions are sufficiently favorable.

Taken together, Figs. 5 and 6 demonstrate that RBOL consistently yields lower GFP and BOP across a wide range of SNRs for both $D=4$ and $D=6$. The gains are most pronounced in the low and intermediate SNR regime (around −3 dB to 0 dB), where the gate and ranking decisions are nontrivial and system performance is most sensitive to prediction errors. In contrast, at very low SNRs, all schemes saturate to outage, while at very high SNRs, all schemes approach the oracle bound. These results confirm that RBOL provides improved robustness to channel quality variations and that its decision-aware training objective remains well aligned with the system-level reliability metric under realistic operating conditions.

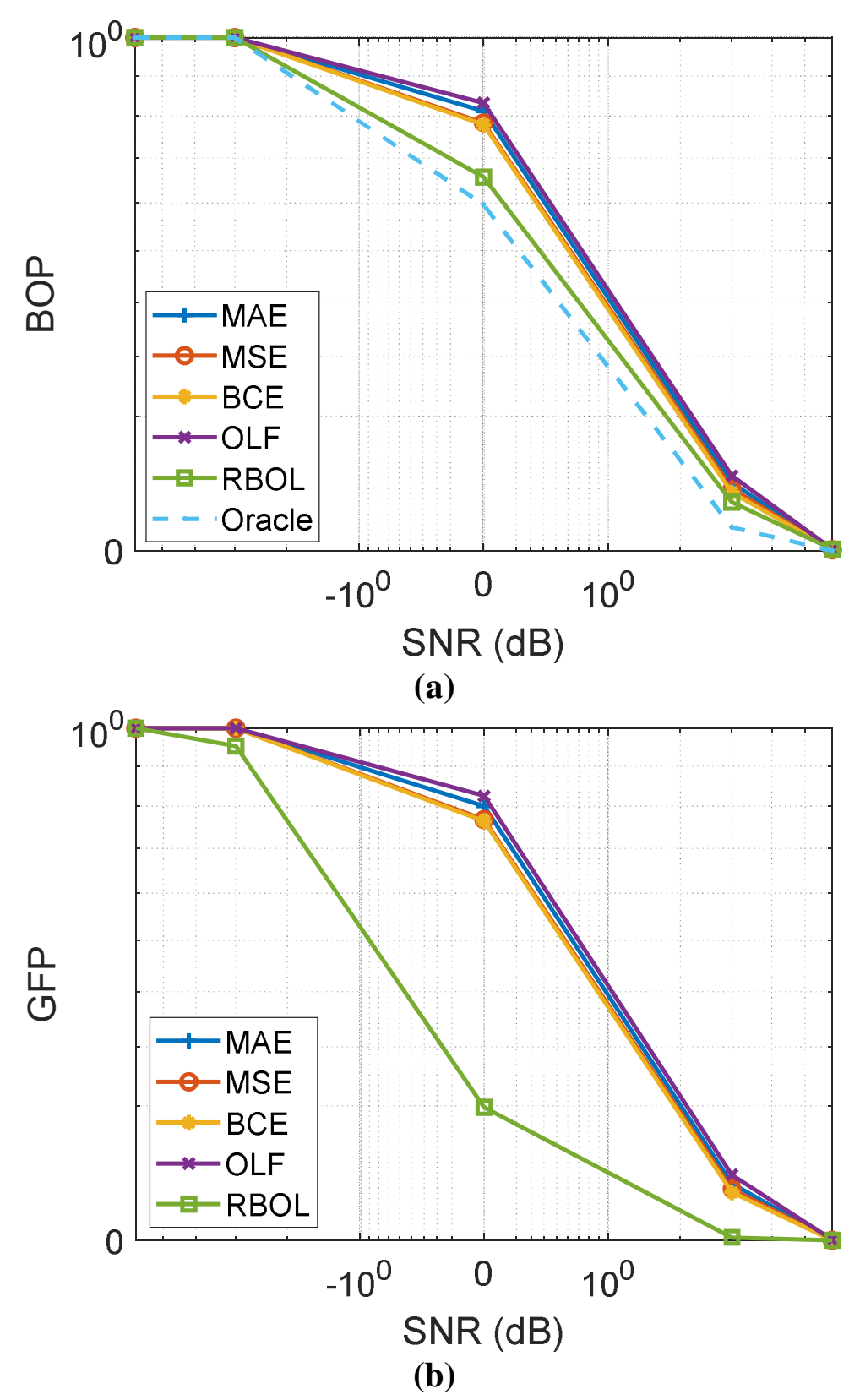

**Fig. 6.** Comparison of different loss functions versus SNR in Example 3 for $D=6$; **(a)** BOP, **(b)** GFP.

*Example 4 (Threshold–Reliability Relationship under GTBA):* In this experiment, we investigate how the gate threshold $q_{\text{th}}$ influences system reliability under the GTBA rule. Unlike Examples 1-3, which fix $q_{\text{th}}$ and vary either $D$ or SNR, here we sweep the gate threshold to expose the relationship between gate failures and selection reliability. All models are trained once under the balanced-stress regime ($\gamma_{\text{th}}=1.2$, $\text{SNR}=0\,\text{dB}$), and evaluation is performed on a shared test set. Results are reported for $D=4$ and $D=6$.

Fig. 7 summarizes the results for $D=4$ using four subfigures: (a) BOP versus $q_{\text{th}}$, (b) GFP versus $q_{\text{th}}$, (c) the BOP-GFP relationship curve, and (d) the average NAR (ANAR), i.e., $\mathbb{E}\left[\left|\mathcal{A}\left(q_{\text{th}}\right)\right|\right]$. From Fig. 7(a), RBOL consistently achieves the lowest BOP across the entire range

of $q_{\text{th}}$. Around the nominal operating point $q_{\text{th}} = 0.4$, RBOL reduces BOP by roughly 26%-43% relative to other losses. While all methods approach the oracle bound for very permissive thresholds, RBOL remains closest to the oracle across the practical operating region. Fig. 7(b) shows that GFP decreases monotonically with increasing $q_{\text{th}}$ for all methods, as expected. For instance, at $q_{\text{th}} = 0.4$, RBOL with a GFP of approximately 0.035 dramatically suppresses gate failures, while OLF, MAE, MSE and BCE remain much higher (with corresponding values of approximately 0.39, 0.34, 0.3 and 0.29, respectively). The curves in Fig. 7(c) confirm that RBOL provides a superior reliability frontier: for a given gate-failure level, RBOL achieves systematically lower BOP than the pointwise losses. This indicates that RBOL improves reliability not merely by relaxing the gate, but by better aligning admission and ranking with the bulk objective. Finally, Fig. 7(d) reports ANAR as a function of $q_{\text{th}}$. It is observed that across a wide range of $q_{\text{th}}$, the proposed RBOL clearly outperforms. For instance, for $q_{\text{th}} = 0.4$, the approximate ANAR values for MAE, MSE, BCE, OLF and RBOL are 4.13, 4.33, 4.36, 3.91 and 6.76, respectively. These results explain the large GFP gap in Fig. 7(b): RBOL admits substantially more candidates (on average), thereby avoiding shortfall ($\left|\mathcal{A}\left(q_{\text{th}}\right)\right| < D$) events, while still achieving the best BOP.

Fig. 8 presents the corresponding results for $D = 6$. Increasing the bulk requirement makes the allocation problem significantly more challenging. As shown in Fig. 8(a), RBOL again yields the lowest BOP over the full $q_{\text{th}}$ sweep. Around $q_{\text{th}} = 0.4$, RBOL reduces BOP by approximately 15%-20% relative to other losses. The oracle curve confirms that these gains are not due to trivial operating points but arise from improved decision quality. Fig. 8(b) reveals that GFP is substantially higher for all methods at $D = 6$, reflecting the difficulty of admitting six reliable resources under balanced stress. Nonetheless, RBOL consistently achieves lower GFP than the pointwise baselines, delaying the onset of frequent gate failures. The curves in Fig. 8(c) further highlight RBOL's advantage: RBOL traces the lower envelope of the BOP-GFP plane, whereas the other losses exhibit inferior tradeoffs, achieving low GFP only at the cost of significantly higher BOP. Fig. 8(d) reports ANAR values. For $D = 6$, OLF, MAE, MSE and BCE admit on average approximately 4 resources near the nominal threshold. RBOL, by contrast, maintains an average admitted set size closer to 7 resources, consistent with its explicit shortfall-aware training objective.

The results of Example 4 demonstrate that sweeping the gate threshold provides a clear and informative view of the reliability relationships induced by different learning objectives. It can also be seen that RBOL consistently achieves a superior tradeoff, validating its design as a decision-aware loss aligned with the GTBA reliability criterion.

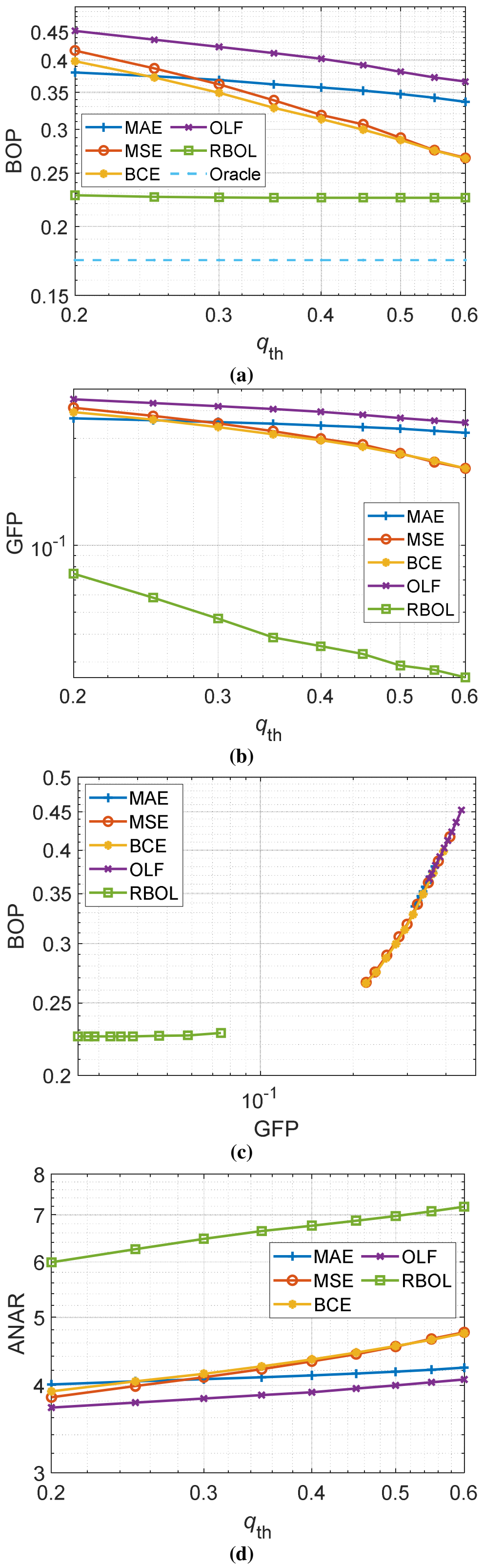

**Fig. 7.** Comparison of loss functions in Example 4 for $D = 4$; **(a)** BOP versus $q_{\text{th}}$, **(b)** GFP versus $q_{\text{th}}$, **(c)** BOP versus GFP, **(d)** ANAR versus $q_{\text{th}}$.

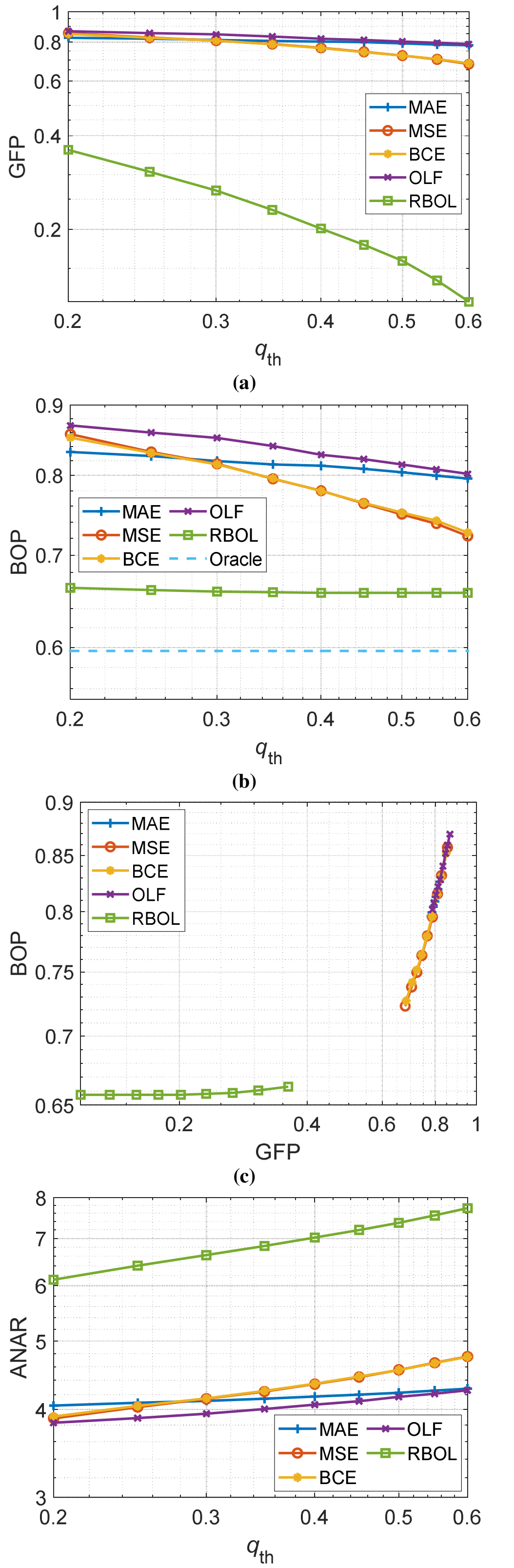

**Fig. 8.** Comparison of loss functions in Example 4 for $D = 6$; **(a)** BOP versus $q_{th}$, **(b)** GFP versus $q_{th}$, **(c)** BOP versus GFP, **(d)** ANAR versus $q_{th}$.

## VI. Conclusion

This paper investigated ML-assisted bulk resource allocation, where a user requires multiple reliable resources simultaneously rather than a single best one. We showed that extending outage-based learning from single-resource to bulk allocation fundamentally changes the nature of the reliability objective: performance is no longer governed by per-resource accuracy alone, but by the collective reliability of a selected set. To address this challenge, we proposed a practical allocation policy, GTBA, and developed a new RBOL that directly targets the bulk outage event induced by this policy.

Through an exact reliability analysis, we decomposed the BOP into gate failures and selection failures, established the OBOP as a strict lower bound, and characterized asymptotic behavior in the large-resource regime. This analysis clarified why conventional pointwise losses (MAE, MSE, BCE and OLF) are poorly aligned with bulk reliability: they do not explicitly control the feasibility of admitting $D$ candidates nor the ranking quality at the top-$D$ boundary. In contrast, RBOL is explicitly designed to suppress both failure modes via a shortfall-aware gate term and a cutoff-aware ranking penalty.

Extensive simulations validated the analysis across multiple operating regimes. In the balanced-stress regime, RBOL reduced the GFP by about an order of magnitude relative to pointwise losses for moderate $D$, and achieved 15%-41% reductions in BOP compared to MAE, MSE, BCE and OLF. Similar trends were observed under light- and heavy-stress regimes, where RBOL consistently remained closest to the oracle bound, while pointwise losses degraded rapidly as the bulk requirement increased. Additional experiments sweeping SNR and gate thresholds further demonstrated that RBOL yields a superior BOP-GFP tradeoff, confirming that its gains are not tied to a specific operating point but arise from better alignment between training and system-level reliability.

Overall, the results showed that bulk reliability cannot be optimized effectively using pointwise objectives, even when combined with ranking-based allocation at inference time. Instead, training objectives must explicitly account for set-level constraints and ranking errors, as achieved by RBOL.

Several directions for future work are promising. One natural extension is multi-user sequential bulk allocation, where multiple users compete for shared resources over time and allocation decisions interact across scheduling instants. Incorporating fairness, inter-user coupling and temporal resource reuse into the bulk outage framework would further bridge the gap between theoretical reliability analysis and practical multi-user systems. Additional extensions include adaptive threshold selection, joint power-resource allocation, and integrating RBOL with more expressive architectures for highly dynamic channels.